\documentclass[reprint, amsmath,amssymb, aps, pra, longbibliography]{revtex4-1}

\usepackage{silence}
\usepackage{soul}
\usepackage{amsmath}   
\usepackage{amssymb}
\usepackage{mathtools}
\usepackage{graphicx}
\usepackage{dcolumn}
\usepackage{bm}
\usepackage{amsfonts}
\usepackage{booktabs}
\usepackage{subfigure}
\usepackage{array}
\usepackage{float}
\usepackage{color}
\usepackage{multirow}
\usepackage{makecell}
\usepackage{amssymb}
\usepackage[colorlinks=true,linkcolor=blue]{hyperref}
\hypersetup{allcolors=blue}
\usepackage[normalem]{ulem}
\usepackage{xcolor}

\newcommand{\ket}[1]{\ensuremath{\left\vert #1 \right\rangle}}
\usepackage{mathtools}

\begin{document}

\preprint{APS/123-QED}

\title{Observation and control of polariton noise in a thermal Rydberg atom electrometer}

\author{Teng Zhang}
    \affiliation{Rydberg Technologies Inc., Ann Arbor, Michigan 48103, USA}
\author{Lu{\'\i}s~Felipe~Gon\c{c}alves}
    \affiliation{Rydberg Technologies Inc., Ann Arbor, Michigan 48103, USA} 
\author{Alisher Duspayev}
    \email{ADuspayev@RydbergTech.com}
    \affiliation{Rydberg Technologies Inc., Ann Arbor, Michigan 48103, USA}   
\author{Georg Raithel}
    \affiliation{Rydberg Technologies Inc., Ann Arbor, Michigan 48103, USA}
\author{David A. Anderson}
    \email{Dave@RydbergTech.com}
    \affiliation{Rydberg Technologies Inc., Ann Arbor, Michigan 48103, USA} 

\date{\today}
             
\begin{abstract} 
The establishment of metrics and control of fundamental noise sources in quantum sensing is a long-standing goal in quantum science and technology. We report on the observation of polariton-mediated optical noise in a thermal Rydberg atom vapor and demonstrate its active control and suppression using a radio-frequency field in Rydberg electrometry. The noise exhibits structured amplification and reduction across the Rydberg electromagnetically induced transparency resonance. This reveals a nonlinear slow-light feedback mechanism in which the polariton medium stores and releases probe-field energy. We show that the reference radio-frequency field used in Rydberg heterodyne receivers can also tune the medium into an operating regime where high electromagnetic-field responsivity coincides with reduced optical noise. This operating-point control enables optimization of the electric-field sensitivity within the constraints imposed by Rydberg-polariton physics. These results establish polariton noise as a previously unrecognized intrinsic noise source in thermal Rydberg vapors and the corresponding control method as a crucial aspect of continued efforts to achieve fundamental sensitivity limits in Rydberg electrometry.
\end{abstract}

\maketitle

\section{Introduction}
\label{intro}

Quantum sensors are an emerging class of technology that can yield unprecedented sensitivities, harness the potential benefits of quantum entanglement and squeezing, and complement classical methods for measurement and metrology~\cite{degenrmp2017}. Prominent examples include atomic magnetometers~\cite{Budker.2007}, optical atomic clocks~\cite{ludlowreview, Fortier:26}, as well as matter-wave interferometers for inertial navigation~\cite{schilipert2025navigation}. Rydberg-atom-based electric-field sensors have emerged as a major platform for quantum electromagnetic sensing~\cite{Anderson.2019,AndersonRadioIEEE.2021,Schlossberger2024}. Exploiting the large electric dipole moments and polarizabilities of atoms in highly-excited Rydberg states~\cite{gallagher},  these quantum systems enable highly sensitive detection and measurement of electromagnetic fields~\cite{Anderson.2014, Fan.2015, Kumar.2016, Meyer.2020, Downes.2020, Kunz.2023}. Rydberg atoms prepared in room-temperature vapor cells enable calibration-free, SI-traceable electrometry~\cite{Holloway2014} and phase- and frequency-sensitive heterodyne (HET) reception of radio-frequency (RF) waves using a reference field~\cite{Anderson.2019,jia2020het} over wide regions of the electromagnetic spectrum~\cite{jau2020, liuMHz2022, daa2020wideband, wu2025THz}. At the device level, these capabilities have been realized in self-calibrated RF electric-field measurement instruments~\cite{AndersonRFMS} and portable Rydberg-atom radio receivers for long-range wireless communications~\cite{AndersonIEEEURSIARx.2024}. Despite this progress, the ultimate sensitivity attainable with such instruments—and the extent to which it is constrained by quantum projection noise~\cite{zhu2024sql, wang2023noise, daa2018whitenoise}, thermal radiation~\cite{whitley2022sql, he2021noise}, and nonlinear atom-field interactions—remain unresolved. 

A particularly important class of nonlinear atom-field interactions arises from the formation and propagation of polaritons under electromagnetically induced transparency (EIT)~\cite{lukin2000polariton, zhu2007noise}. Following its discovery~\cite{harris1990eit}, EIT has been observed and studied in a variety of systems~\cite{fleishauer2005review} and has played important roles in quantum information processing~\cite{duan2020communication}, nonlinear optics~\cite{eit2024nonlinear, eit2024optics}, and metrology~\cite{eit2023metrology}. As a characteristic of EIT, the optical properties of the media for a weak probe field, which near-resonantly drives the transition between the ground and intermediate states, are greatly altered by an additional strong laser field, commonly referred to as the coupler, which resonantly drives the transition between the intermediate and target metastable states. In the context of this study, the latter is a Rydberg state. The intermediate state rapidly decays to the ground state via natural decay. One hallmark of EIT is the reduced absorption of the probe field at the two-photon resonance. Moreover, the medium exhibits a large positive linear dispersion within this transparency window, which results in a reduction of the group velocity. This gives rise to a phenomenon often referred to as ``slow light''~\cite{xiao2011review}. When passing through the medium, a probe photon converts into a coherent superposition of a photon and a single, distributed, coherent atomic excitation between the ground and target states. The resulting quasiparticles, known as polaritons, propagate at reduced group velocity~\cite{fleischauer2004reduced} and can exhibit strong nonlinearities, correlations, and quantum phases~\cite{zhu2020polariton, icfo2016polariton}. Despite extensive study in other systems~\cite{lukin2008photon}, the noise generated by polariton dynamics in dilute thermal vapors, and its consequences for Rydberg electrometry, have remained largely unexplored. Here, we directly observe and characterize polariton-mediated optical noise in a thermal Rydberg vapor, distinguish it from conventional optical noise such as photon shot noise, and show that it can impose a fundamental constraint on attainable electric-field sensitivity. We further demonstrate active control and suppression of this noise using a resonant reference radio-frequency field while maintaining high electromagnetic-field responsivity.

\begin{figure*}[t!]
\centering
\includegraphics[width=1\linewidth]{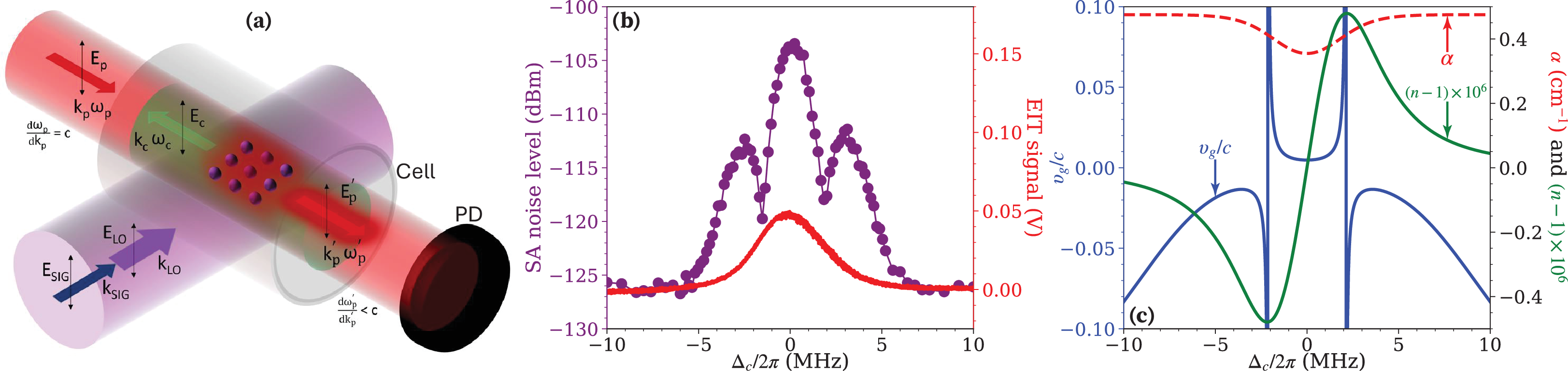}
\caption{ 
(a) Conceptual diagram of the experimental setup, including the probe beam (red), the coupler beam (green), RF fields (purple), and their wave and polarization vectors. Primed probe-beam quantities are inside the Rydberg vapor. A photodiode (PD) measures the probe power transmitted through the EIT medium. (b) Measured background-subtracted probe-light transmission  (red) and polariton noise (purple) at $\Omega_c/(2\pi)\approx0.4$~MHz versus coupler beam detuning $\Delta_c$. The utilized Rydberg state is $\ket{42D_{5/2}}$. (c) Theoretical probe-light absorption $\alpha$ (dashed red), scaled refractive index (n-1)$\times$10$^6$ (solid green), and group velocity $v_g$ in units of $c$ (solid blue) in the Rydberg medium.}
\label{fig:setup}
\end{figure*}

Our paper is organized as follows. In Sec.~\ref{exp}, the utilized experimental setup is described. In Sec.~\ref{measurement}, we present experimental results that elucidate the nature and characteristics of the polariton-induced noise. In Sec.~\ref{theory}, we develop a theoretical EIT model that describes polariton propagation in the room-temperature vapor medium. The physical interpretation is based upon feedback between probe-laser power and polariton speed, which can buffer power fluctuations by leveraging the dynamics of energy storage in the atomic medium. In Sec.~\ref{noise_tuning}, we demonstrate tuning of the polariton noise using a resonant RF field, allowing us to achieve an improved sensitivity of the Rydberg HET RF sensor. In Sec.~\ref{discussion}, findings and implications of our study are summarized.

\section{Experimental setup}
\label{exp}

The major part of the utilized setup is depicted in Fig.~\ref{fig:setup}~(a). The experiment is performed using a room-temperature vapor of $^{133}$Cs atoms in a 30-mm-long cylindrical glass cell of 10-mm diameter. A probe laser at 852~nm (red) is frequency-stabilized at a detuning $\Delta_p/2\pi\approx+250$~MHz relative to the $\ket{6S_{1/2}, F=4}\rightarrow\ket{6P_{3/2}, F'=5}$ transition for zero-velocity atoms. Probe light passing through the vapor cell is filtered using a narrowband optical filter [not shown in Fig.~\ref{fig:setup}~(a)] and collected by a fast photodiode (PD). The PD photocurrent is then amplified by a 3.5-MHz-bandwidth transimpedance amplifier (TIA), the output of which is recorded using either an analog-to-digital converter (ADC) for time-domain analysis or a spectrum analyzer (SA) for frequency analysis. A counter-propagating coupler laser at 511~nm (green) is overlapped with the probe beam in the vapor cell using a dichroic mirror. The polarizations of both lasers are linear and parallel to each other. The probe and coupler beams have powers of 1.1~mW and 340~mW and $1/e^2$ radii ($w_0$) of 10~mm and 11~mm, respectively. The corresponding Rabi frequencies at the centers of the respective beams are $\Omega_p/(2\pi)=$~1.8~MHz and $\Omega_c/(2\pi)=$~0.4~MHz. 

To isolate the atomic physics that gives rise to the polariton-induced optical noise, laser frequency drifts, jitter, and phase noise must be minimized. Therefore, the probe and coupler lasers are stabilized to an ultra-stable optical frequency comb using optical phase-locked loops, as described in detail in~\cite{remy2024receiver}, leading to respective laser linewidths well below 100~Hz. This results in relative frequency stabilities of both lasers of $\lesssim 2 \times 10^{-13}$. The coupler detuning, $\Delta_c$, from the $\ket{6P_{3/2}, F'=5}\rightarrow\ket{42D_{5/2}}$ and $\ket{6P_{3/2}, F'=5}\rightarrow\ket{43S_{1/2}}$ transitions used in this work is scanned via the frequency offset lock of the coupler laser to a comb tooth~\cite{remy2024receiver}. The transmission of the probe light is recorded as a function of $\Delta_c$ to obtain EIT spectra. Spectral analysis of the transmitted probe light yields the noise spectra discussed in our paper.

\section{Observation of polariton-induced optical noise}
\label{measurement}

For polariton noise measurement, we separately record and process the dc and ac voltages from the TIA, which amplifies the PD signal, as a function of coupler detuning, $\Delta_c$. The dc component is a measure of the overall light absorption, which exhibits an EIT absorption dip [red curve in Fig.~\ref{fig:setup}~(b)]. A typical full width at half maximum of the EIT line in the present work is $\Delta\nu_{\mathrm{EIT}}~\approx$~4.5~MHz. The ac component is processed by the SA, yielding the noise spectrum [purple curve in Fig.~\ref{fig:setup}~(b)]. Here, we focus on the noise level as a function of $\Delta_c$ in the vicinity of 10~kHz, the intermediate frequency (IF) used for HET reception in Sec.~\ref{noise_tuning}.

In Fig.~\ref{fig:setup}~(b), we see that the EIT resonance is accompanied by excess noise above the generic optical and atomic noise of the system ($\approx-126$~dBm). This excess occurs only in the presence of the EIT resonance, indicating that it is generated or strongly enhanced by the atom-light interaction under EIT conditions. We therefore refer to this contribution as polariton-induced optical noise. As seen in Fig.~\ref{fig:setup}~(b), it peaks at about 25~dB above the system's noise floor at the center of the EIT line. Close to the inflection points of the EIT line at $\Delta_c\approx\pm\Delta\nu_{\mathrm{EIT}}/2$, the polariton noise dips to only about 6~dB above the noise floor. The noise dips are surprisingly narrow; the observed width of the dips is only about $\Delta\nu_{\mathrm{EIT}}/10$. In the wings of the EIT line, the polariton noise exhibits secondary maxima about 15~dB above the noise floor. Both the EIT spectrum and the noise curve in Fig.~\ref{fig:setup}~(b) are symmetric about $\Delta_c=0$, and the overall widths of the EIT and the polariton noise features are similar.

The microscopic origin of the polariton noise may lie in effective interactions between polaritons~\cite{sun2017direct, chafatinos2020polariton}, fluctuations in the collective atomic excitation shared by the polariton field~\cite{weidemuller2012polariton}, and dynamical variations of macroscopic medium parameters such as refractive index, absorption, and group velocity.  Development of a complete quantum model is beyond the scope of the present work. Instead, we formulate a phenomenological slow-light feedback picture that captures the observed regimes of noise amplification and reduction.

In Fig.~\ref{fig:setup}~(c), we plot the probe absorption coefficient $\alpha$ (dashed red), the probe refractive index, $(n-1)$ (solid green), and the polariton group velocity in the EIT medium, $v_g$ (solid blue), as a function of $\Delta_c$. The latter is obtained using

\begin{equation}
    v_g = \frac{c}{n (\nu_p)+ \nu_p \frac{dn(\nu_p)}{d \nu_p}} \quad.
\label{eq:vg}    
\end{equation}

The calculations are done for the probe and coupler Rabi frequencies used in the experiment (see Sec.~\ref{exp}). The utilized model is described in detail in Sec.~\ref{theory}. Near the center of the EIT line at $\Delta_c = 0$, it is typically $v_g / c < 0.05$. Due to the underlying behavior of $dn(\nu_p)/d \nu_p$, the value of $v_g/c$ gradually rises with increasing $| \Delta_c |$, and it diverges at poles located at $\pm \Delta_{cp} \approx \pm 2 \pi \times 2.2~$MHz. 
%At the poles $dn(\nu_p)/d \nu_p \approx 0$. 
The ``slow-light'' region found at the center of the EIT line~\cite{scully1999slow, lukin2001controlling} is delimited by these poles. Outside the poles, {\sl{i.e.}}, for $|\Delta_c| > \Delta_{cp}$, the value of $v_g$ becomes negative.

\begin{figure} [t!]
\centering
\includegraphics[width=1.\linewidth]{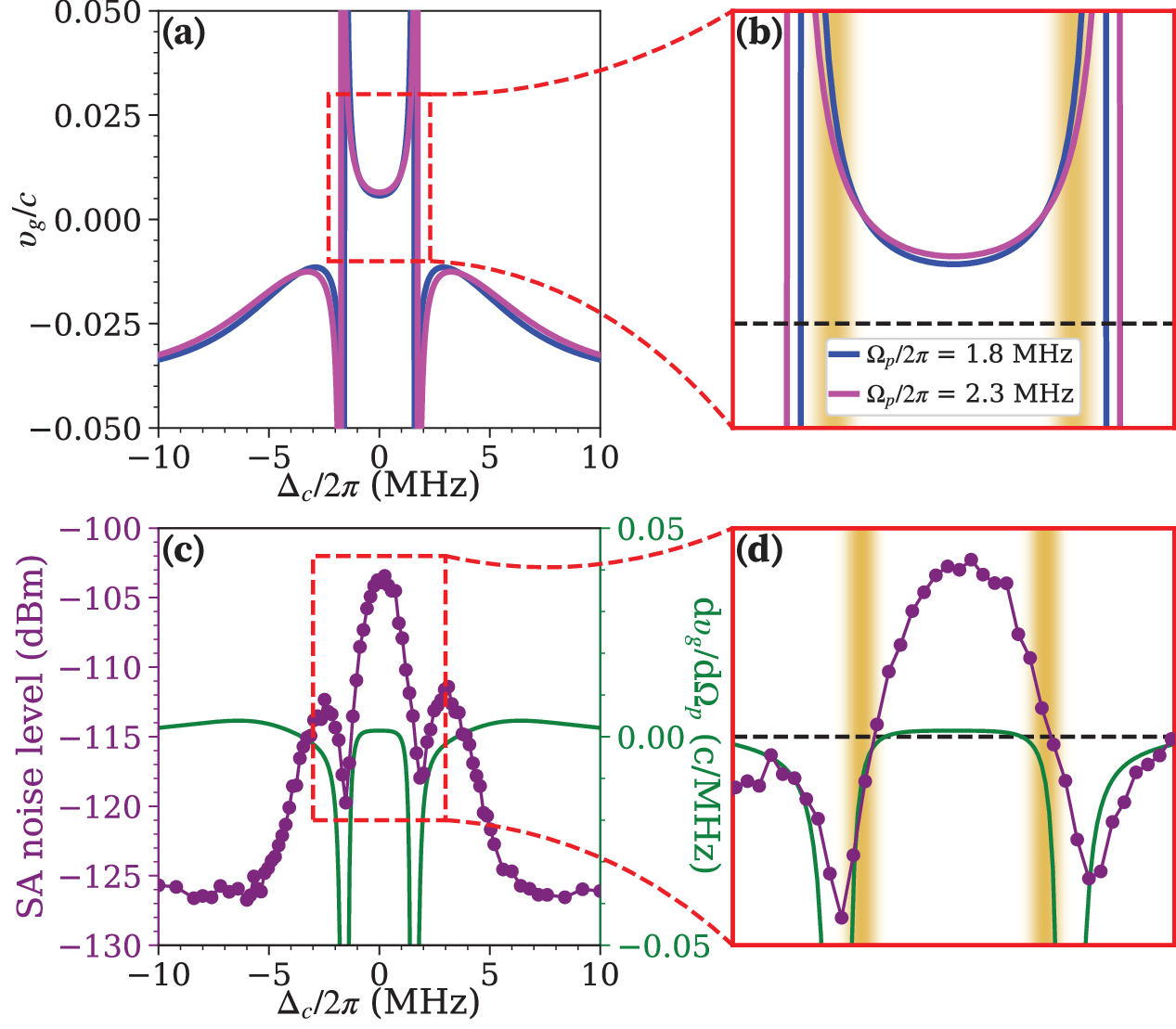}
\caption{Polariton noise amplification and reduction near EIT resonance. (a) Calculated polariton group velocity, $v_g$, versus coupler detuning, $\Delta_c$, for the probe Rabi frequencies, $\Omega_p$, indicated in panel (b). The latter is a zoom-in of panel (a). (c) $dv_g/d\Omega_p$ versus $\Delta_c$ (green) overlaid with the measured polariton noise level (purple). Panel (d) is a zoom-in of panel (c). Horizontal dashed black line in (b) and (d) corresponds to $v_g = 0$ and $d v_g/ d \Omega_p = 0$, respectively. In (c) and (d), the orange shading indicates the regions where $v_g > 0$ and $dv_g/d\Omega_p < 0$. There, our model predicts polariton noise reduction through a negative-feedback mechanism (see text for details).}
\label{fig:exp}
\end{figure}

To gain further insight, in Figs.~\ref{fig:exp}~(a) and~(b) we show $v_g$ as a function of $\Delta_c$ for two probe Rabi frequencies $\Omega_p /(2\pi)=$~1.8~MHz (blue) and 2.3~MHz (magenta). Near the center of the EIT resonance, a lower (larger) $\Omega_p$ results in a reduced (increased) $v_g$, {\sl{i.e.}}, $dv_g/d\Omega_p>0$. At coupler detunings of $\Delta_c = \pm \Delta_{c0} \approx \pm 2 \pi \times 1.5$~MHz, the two curves cross, {\sl{i.e.}}, $dv_g/d\Omega_p \approx 0$ at these detunings. There are two symmetric windows between $\Delta_{c0} < \Delta_c < \Delta_{cp}$ and $-\Delta_{cp} < \Delta_c < - \Delta_{c0}$, where $v_g > 0$ and $dv_g/d\Omega_p < 0$. 

In Figs.~\ref{fig:exp}~(c) and~(d), we plot the computed $dv_g/d\Omega_p$-curve (solid green line) overlaid with the same experimental SA noise spectrum (connected purple dots) as in Fig.~\ref{fig:setup}~(b). There, the aforementioned windows in which $v_g > 0$ and $dv_g/d\Omega_p < 0$ (orange shading) coincide fairly closely with the regions where the experimentally observed polariton noise is reduced. We attribute minor discrepancies to the Rabi-frequency choices in the calculation and optical field inhomogeneities in the experiment. Interestingly, to find a match between the calculations and the experimental observations, the coupler Rabi frequency in the former had to be set to $\Omega_c/(2\pi)=$~1.5~MHz, almost a factor of 4 larger than the experimental counterpart. To explain that, we note that the excitation of Rydberg polaritons occurs at an effective Rabi frequency $\Omega_{eff}=\sqrt{N}\Omega_1$, where $N$ is the number of atoms that share one polariton quantum excitation, and $\Omega_1$ is the single-atom Rabi frequency~\cite{Dudin2012, dodel2020}. For a room-temperature Cs vapor density and a polariton radius of around 10~$\mu$m at the selected Rydberg state, we estimate $N\approx$~10 in our case, where we account for the reduction in effective atom density due to the Doppler effect. This leads to $\Omega_{eff}/(2\pi)\approx$~1.3~MHz, in much better agreement between theory and experiment. The observations in Fig.~\ref{fig:exp} form a basis for the feedback mechanism described next.

\begin{figure*}[t!]
\centering
\includegraphics[width=0.9\linewidth]{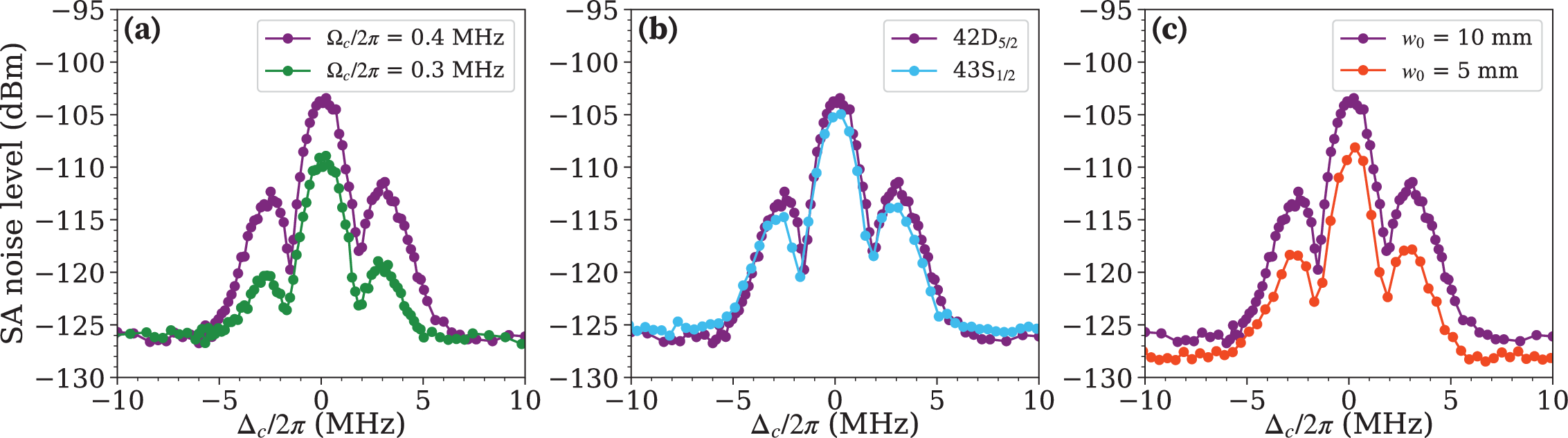}
\caption{Rydberg polariton noise versus coupler detuning, $\Delta_c$, at different coupler Rabi frequencies, $\Omega_c$, (a), using different Rydberg levels (b), and with different EIT beam volumes (c). In the latter panel, the indicated $w_0$~=~10~mm (5~mm) corresponds to the EIT beam volume of 3.0~cm$^{3}$ (0.75~cm$^{3}$). $w_0$~=~10~mm is also used for the data in (a) and (b). In (a) and (c), the utilized Rydberg state is $\ket{42D_{5/2}}$. In (b), the coupler Rabi frequencies are slightly different ($\Omega_c/(2\pi)=$~0.3~MHz (0.4~MHz) for the $\ket{43S_{1/2}}$ ($\ket{42D_{5/2}}$) state) due to available coupler power and a difference in the respective electric-dipole transition matrix elements. See text for discussion.}
\label{fig:compare}
\end{figure*}

To leading order, the optical energy associated
with the probe field and stored in the EIT medium is $W = P_p L / v_g$, where $P_p$ is the probe power and $L$ is the medium length. Therefore, the medium can act as a dynamic reservoir that can buffer energy variations of the probe beam as it travels through the medium. In the $\Delta_c$-ranges in which $v_g > 0$ and $dv_g/d\Omega_p < 0$, an increase in $P_p$ (and, hence, $\Omega_p$) results in a reduced $v_g$, {\sl{i.e.}}, the EIT medium will be able to store an excess probe energy of up to $dW = P_p L |dv_g| /v_g^2$. The excess stored energy corresponds to a reduction in the probe beam's power transmitted through the medium. Vice versa, a decrease in probe power results in a larger $v_g$, causing the EIT medium to release energy of up to $dW = P_p L |dv_g| /v_g^2$. The latter is then added to the probe beam's output. In any case, the noise in probe-beam power remains small after passage through the medium. The described negative feedback requires $v_g > 0$ and $dv_g/d\Omega_p < 0$ as conditions for polariton noise reduction. A minimum of polariton-induced noise occurs if a fluctuation in probe power changes $v_g$ and the energy stored in the EIT medium such that the fluctuation is canceled.

Conversely, if $v_g > 0$ and $dv_g/d\Omega_p > 0$, probe noise is amplified by the EIT medium (positive-feedback process). As seen in Fig.~\ref{fig:exp}, this case occurs near the center of the EIT line, $|\Delta_c| \approx 0$. Moreover, it can be seen in Fig.~\ref{fig:exp} that $v_g < 0$, found when $|\Delta_c| \gtrsim \Delta_{cp}$, also causes an enhanced polariton noise. Negative $v_g$ is generally associated with anomalous dispersion effects, including superluminal pulse propagation through media~\cite{Wang2000} and reverse wave-packet motion in the medium~\cite{ye2015observation}. Therefore, the simple positive-slow light buffering picture no longer applies. The present observations imply that the anomalous-dispersion regime does not provide the same stabilizing feedback as under conditions $v_g > 0$ and $dv_g/d\Omega_p < 0$.

In Fig.~\ref{fig:compare}, we show that the measured polariton noise exhibits similar characteristics over a range of coupler Rabi frequencies, $\Omega_c$,~(a), Rydberg quantum states~(b), and EIT beam volumes varied by changing beam radii, $w_0$, (c). The polariton noise ``dips" and central lobes, where noise is enhanced, are consistently located in the negative- and positive-feedback domains, respectively. The side lobes, where the noise is also amplified, are in negative-$v_g$ domains. In Fig.~\ref{fig:compare}~(a), we see that larger $\Omega_c$ results in a higher noise level and a larger relative depth of the noise dips. In Fig.~\ref{fig:compare}~(b), the polariton noise for two Rydberg states with different angular momentum but equivalent $\Omega_c$ shows nearly identical features. This observation suggests that the polariton noise is mainly determined by the EIT coupling strengths, not by the type of Rydberg states. Finally, in Fig.~\ref{fig:compare}~(c), we use different optical beam sizes while maintaining the same Rabi frequencies at the beam centers for both the probe and coupler lasers. The larger beams and interaction volumes result in higher noise levels and generally sharper features in the polariton noise curves. This is also expected because larger interaction volumes involve more atoms to contribute to the noise.

\section{EIT response model and slow-light feedback}
\label{theory}

The model used here does not directly calculate
quantum-noise spectra or microscopic fluctuation correlations. Instead, it computes the steady-state optical response of the Rydberg-EIT medium, including absorption, refractive index, group velocity, and their dependence on optical and RF Rabi frequencies. These quantities are then used to identify operating regimes in which slow-light feedback is expected to amplify or reduce transmitted-probe intensity noise. In the theoretical model used for Figs.~\ref{fig:setup} and~\ref{fig:exp} and in the remainder of this paper, we employ the density matrix formalism. The Hilbert space we use here, $\{ \vert i \rangle, i = 1,2,3,4 \}$, is spanned by the ground state, $\ket{1} = \ket{6S_{1/2}}$, the intermediate state, $\ket{2} = \ket{6P_{3/2}}$, with $\Gamma_{12}/2\pi=$~5.2~MHz, and a set of up to two Rydberg states. The dynamics of the system is calculated using the Lindblad equation
\begin{equation}
    \dot{\hat{\rho}} = -\frac{i}{\hbar} [\hat{H}, \hat{\rho}] + L(\hat{\rho}) \quad,
\label{eq:rho}
\end{equation}
where $\hbar$ is the reduced Planck constant. The diagonal elements of the Hamiltonian $\hat{H}$ are given by the dressed energy levels of the system, $H_{11} = 0$, $H_{22} = -\hbar(\Delta_p - k_p v)$,  $H_{33} = H_{22}-\hbar(\Delta_c + k_c v)$, $H_{44} = H_{33} - \hbar\Delta_{RF}$, where the wavenumbers for probe and coupler, $k_p$ and $k_c$, are defined positive, $v$ is the atom velocity in probe beam direction, and $\Delta_{RF}$ is the detuning of the Rydberg transition, if there is a second coupled Rydberg state. The Doppler shift of the Rydberg transition is neglected because this is a microwave transition that exhibits shifts in the range of tens of kHz, too small to be relevant in the present work. The off-diagonal elements are $H_{ij}=\frac{1}{2}\hbar\Omega_{ij}$ for $i = j \pm 1$. The Rabi frequencies are $\Omega_{ij}={\bf{E}}_{ij} \cdot \langle i \vert \hat {\bf{d}} \vert j \rangle /\hbar$,  with field amplitudes ${\bf{E}}_{ij}$ and dipole operator $\hat{\bf{d}}$. The Lindblad operator contains diagonal terms with transition rates $L_{ii}=-(\sum_{j\neq i}\Gamma_{ji}) \rho_{ii}  +\sum_{j \neq i}(\Gamma_{ij}\rho_{jj})$, and off-diagonal terms with level dephasing rates $L_{ij}=-\frac{1}{2} [ (\sum_{k\neq i}\Gamma_{ki}) + (\sum_{k\neq j}\Gamma_{kj} ) +\gamma_i+\gamma_j ] \rho_{ij}$, where $\Gamma_{ki}$ denotes the electromagnetic decay rate from $\vert i \rangle$ to $\vert k \rangle$ and $\gamma_i$ denotes the collisional dephasing rate of $\vert i \rangle$. We further account for a global atom recycling rate, for which we use 2$\pi$ times the inverse of the average atom-field interaction time.  

The atoms have a thermal Maxwell distribution, where, in the one-dimensional case, $P(v)=\frac{1}{\sqrt{2\pi}\sigma_v}exp(-\frac{v^2}{2\sigma^2_v})$, with $\sigma_v=\sqrt{\frac{k_B T}{M}}$ ($k_B$ is the Boltzmann constant, $T$ is the temperature, and $M$ is the atomic mass of $^{133}$Cs atom). The absorption coefficient is calculated by the velocity average of the steady-state density matrix element $\rho_{21}(v,\Delta_c)$, $\alpha(\Delta_c) = \frac{2n_V d_{21}}{\epsilon_0 E_{21}}\int {\rm{Im}} [\rho_{21}(v,\Delta_c)]P(v)dv$. The index of refraction of the sample can be calculated by $(n-1)= \frac{\omega_{21}}{2c}\frac{2n_V d_{21}}{\epsilon_0 E_{21}}\int {\rm{Re}} [\rho_{21}(v, \Delta_c)] P(v)dv$. 
There, $d_{21}=1.99 ea_0$ is the isotropic dipole moment of the probe transition. Note that, in general, $\rho_{21}(v, \Delta_c, \Omega_{RF})$ depends on all parameters of the system. Here we only list dependence on $v$, $\Delta_c$, and $\Omega_{RF}$ as they are of interest in this calculation. The group velocity of the probe light is calculated using Eq.~\ref{eq:vg}.

The model provides a complete picture of the atoms' optical response as a function of all variables, including the detunings and Rabi frequencies of all fields. It omits the spatial structure of the fields, the magnetic substructure of the atoms, and atom-atom interactions. Yet, the present model is sufficient to explain the measured polariton noise behavior as a function of coupler detuning, $\Delta_c$, and coupler and RF Rabi frequencies, $\Omega_{c}$ and $\Omega_{RF}$.

\begin{figure}[t!]
\centering
\includegraphics[width=1\linewidth]{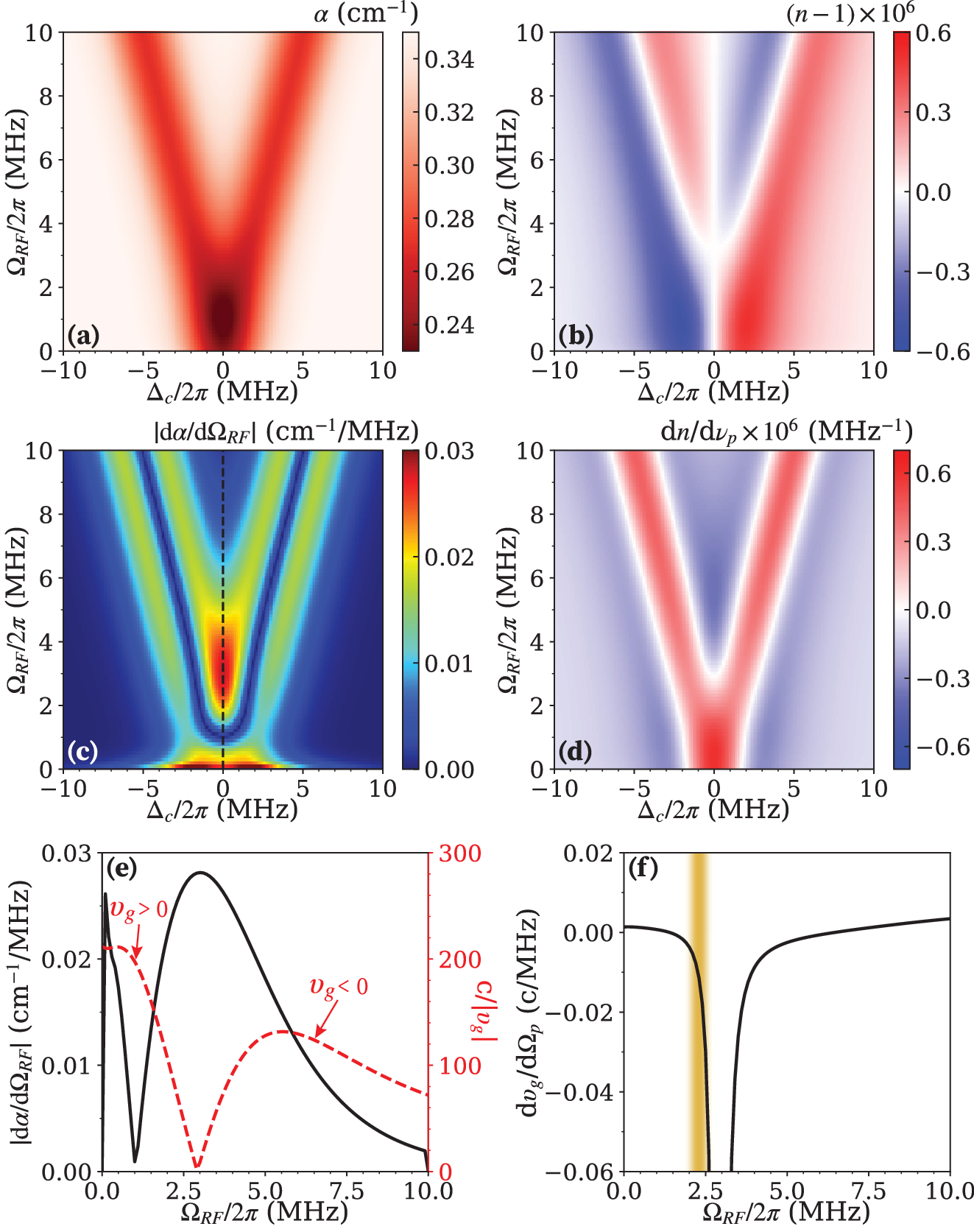}
\caption{Theoretical model.
(a) Probe light absorption coefficient, $\alpha$, as a function of coupler detuning $\Delta_c$ and Rabi frequency of the RF field, $\Omega_{RF}$. 
(b) Rescaled probe refractive index, $10^6\times[n(\Delta_c, \omega_{RF})-1]$. 
(c) Responsivity $|[d\alpha/d\Omega_{RF}](\Delta_c, \omega_{RF})|$. 
(d)  Derivative of the probe refractive index with respect to the probe frequency, $[dn/d\nu_p] (\Delta_c, \omega_{RF})$. 
(e) Responsivity $|d\alpha/d\Omega_{RF}|$ (solid black curve) and $c/|v_g|$ (dashed red curve) as a function of $\Omega_{RF}$ for fixed $\Delta_c=$~0. Branches with $v_g >0$ and $v_g <0$ are indicated on the $c/|v_g|$-curve.
(f) Feedback figure $d v_g/ d \Omega_p$ as a function of $\Omega_{RF}$ for fixed $\Delta_c=$~0. The orange shading indicates the region where the theoretical model predicts polariton noise reductions via negative feedback (see text for details).}

\label{fig:figureTh1}
\end{figure}

Figure~\ref{fig:figureTh1} shows calculated EIT response functions for a room-temperature $^{133}$Cs Rydberg vapor under conditions representative of the experiment. Resonant Autler-Townes splitting, seen on the $\alpha (\Delta_c, \Omega_{RF})$-map [Fig.~\ref{fig:figureTh1}~(a)], underlies numerous applications of Rydberg-EIT in calibration-free electric field sensing and RF reception~\cite{Holloway2014, jia2020het}. The derivative, $|d\alpha/d\Omega_{RF}|$ [map in Fig.~\ref{fig:figureTh1}~(c)], is a metric for the responsivity of Rydberg HET RF receivers. The best operating point on this map, where $|d\alpha/d\Omega_{RF}|$ is largest, occurs at $\Delta_c = 0$ and $\Omega_{RF}/(2\pi) \approx 3$~MHz. EIT is closely intertwined with slow-light propagation. The probe's group velocity, $v_g$, which is equivalent to the polariton propagation speed, is given by Eq.~\ref{eq:vg}, where $n$ is the medium refractive index for the probe light and $dn/d\nu_p$ is its derivative with respect to the probe frequency, $\nu_p$. The $n(\Delta_c, \Omega_{RF})$-and $dn/d\nu_p (\Delta_c, \Omega_{RF})$-maps [Figs.~\ref{fig:figureTh1}~(b) and~(d), respectively] readily yield $v_g$. Slow Rydberg polariton propagation, {\sl{i.e.}}, $v_g \ll c$ with $v_g >0$, requires $dn/d\nu_p$ to be large and positive. Slow-polariton windows occur in selected regions on the $(\Delta_c, \Omega_{RF})$-plane and typically cover probe frequency bandwidths spanning $\lesssim 10$~MHz.

For polariton noise suppression via the negative feedback described in Sec.~\ref{measurement}, we require that (i) $d v_g/ d \Omega_p < 0$, (ii) $v_g >0$ for stable polariton propagation, and (iii) $v_g \lesssim  c/50$ for high buffering capacity for polariton-field energy.  Under the conditions assumed in Fig.~\ref{fig:figureTh1}, these conditions are satisfied at $\vert \Delta_c/(2 \pi) \vert \approx $~2~MHz, when $\Omega_{RF} = 0$. This is seen in Fig.~\ref{fig:exp} above, where there is no RF field applied. 

We use the model to identify how a resonant reference RF field that couples two Rydberg states (with the corresponding Rabi frequency $\Omega_{RF}$) can actively control the polariton-noise landscape by altering both the RF responsivity and the slow-light feedback condition. In Fig.~\ref{fig:figureTh1}~(e), where the responsivity metric $\vert d \alpha / d \Omega_{RF} \vert $ and $c/|v_g|$ versus $\Omega_{RF}$ along the cut $\Delta_c=0$ are plotted, we see that large responsivity and slow polariton propagation coincide at $\Omega_{RF}/(2\pi) \approx 2.8~$MHz. Figures~\ref{fig:figureTh1}~(e) and~(f) show that in the range 2~MHz~$ \lesssim \Omega_{RF}/(2\pi) \lesssim $~2.8~MHz and $\Delta_c=0$, there is a window in which $0 < v_g \lesssim c/50$ and $d v_g/ d \Omega_p < 0$. That is, the conditions for the polariton noise suppression via negative feedback are satisfied, while the responsivity metric $\vert d \alpha / d \Omega_{RF}\vert $ is near its peak. This results in a maximum signal-to-noise ratio (SNR) of Rydberg-HET receivers at $\Omega_{RF}/(2\pi) \approx 2.5$~MHz and $\Delta_c = 0$, under the conditions assumed in Fig.~\ref{fig:figureTh1}. In the following, we investigate this domain experimentally.

\section{Active suppression of polariton noise using a resonant RF field}
\label{noise_tuning}

To demonstrate improved RF sensitivity and noise tuning by an additional RF field, we implement RF detection via the Rydberg HET method~\cite{jia2020het}. For that, two linearly polarized microwave fields at $\sim$~9.937~GHz, combined and emitted from the same horn, are shone onto the vapor cell. The horn is located at a distance of approximately 18~cm from the cell. One of the fields serves as a reference local oscillator (LO) field, and the other as a signal (SIG) field. The SIG frequency, $f_{SIG}$, differs from the reference LO frequency, $f_{LO}$, by a small difference that defines the intermediate frequency, $f_{IF} = f_{SIG} - f_{LO}$, for receiver operation. Here, we use $f_{IF}=$~10~kHz. In this configuration, the atoms respond to the beat of the net field, which consists of the interfering SIG and LO fields. This induces amplitude modulation of the absorption coefficient, $\alpha$, of the medium at the frequency $f_{IF}$. This modulation is observed in the optical readout of the transmitted probe laser using the SA. The SA reading at $f_{IF}$ constitutes the output power, $P_{out}$, of the Rydberg atomic receiver. In an ideal case, $P_{out}$ is proportional to $|d\alpha /d\Omega_{RF}|^2$ and thus depends on $\Omega_{RF}$. Overall, the HET method is conducive to sensitive RF reception~\cite{jia2020het, remy2024receiver} and suitable for characterization of nonlinearities occurring in the atomic receivers~\cite{luis2024nonlinearity}.

We now focus on the SA noise level at $f_{IF}$. First, with the SIG turned off, we find the net noise level of the Rydberg-HET receiver. It consists of a technical contribution (residual laser frequency and intensity noise and noise in utilized electronic parts), optical shot noise from the probe light, and the polariton noise introduced by the nonlinear atom-light interaction in the EIT medium. These noise sources are discerned by selectively removing certain ones while retaining others. The results of this procedure are summarized in Table~\ref{tab:tab1}. It can be seen that the optical noise of the probe field and atom-number noise raised the noise floor by about 6~dB (``coupler off" in Table~\ref{tab:tab1}), while the polariton noise, induced by turning on the coupler, added another 23~dB. Thus, the latter arises as an important fundamental limitation for Rydberg-HET RF receivers. 

\begin{table}[!h]
\begin{center}
\small
    \begin{tabular}{lc} \toprule
     %\hline
     \textbf{Condition} & \textbf{SA signal (dBm)} \\ [0.5ex] \midrule
     \hline
    SA noise (instrument baseline) & -132 \\ \hline
    $\Omega_c = \Omega_p = 0$ (electronics noise) & -131.5  \\   \hline
    $\Omega_c/(2\pi)=0.4$~MHz, $\Omega_p = 0$ (probe off) & -131.5  \\ \hline
    $\Omega_c =0 $, $\Omega_p/(2\pi)=1.8$~MHz  (coupler off) & -126 \\   \hline
     \makecell[l]{ 
                $\Omega_c/(2\pi) =0.4$~MHz, \\ 
                $\Omega_p/(2\pi)=1.8$~MHz, \hspace{20pt}(EIT conditions)   \\  
                $\Delta_c=0$} & 
      -103 \\ [0.5ex] 
      \hline
     \bottomrule
    \end{tabular}
\caption{\label{tab:tab1}Rydberg-HET noise levels at $f_{IF}$ under selected conditions for the SA resolution bandwidth of 1~Hz. In the first row of data, the amplified PD signal is not connected to the SA (in all other rows it is).}
\end{center}
\end{table}

\begin{figure}
\centering
\includegraphics[width=1\linewidth]{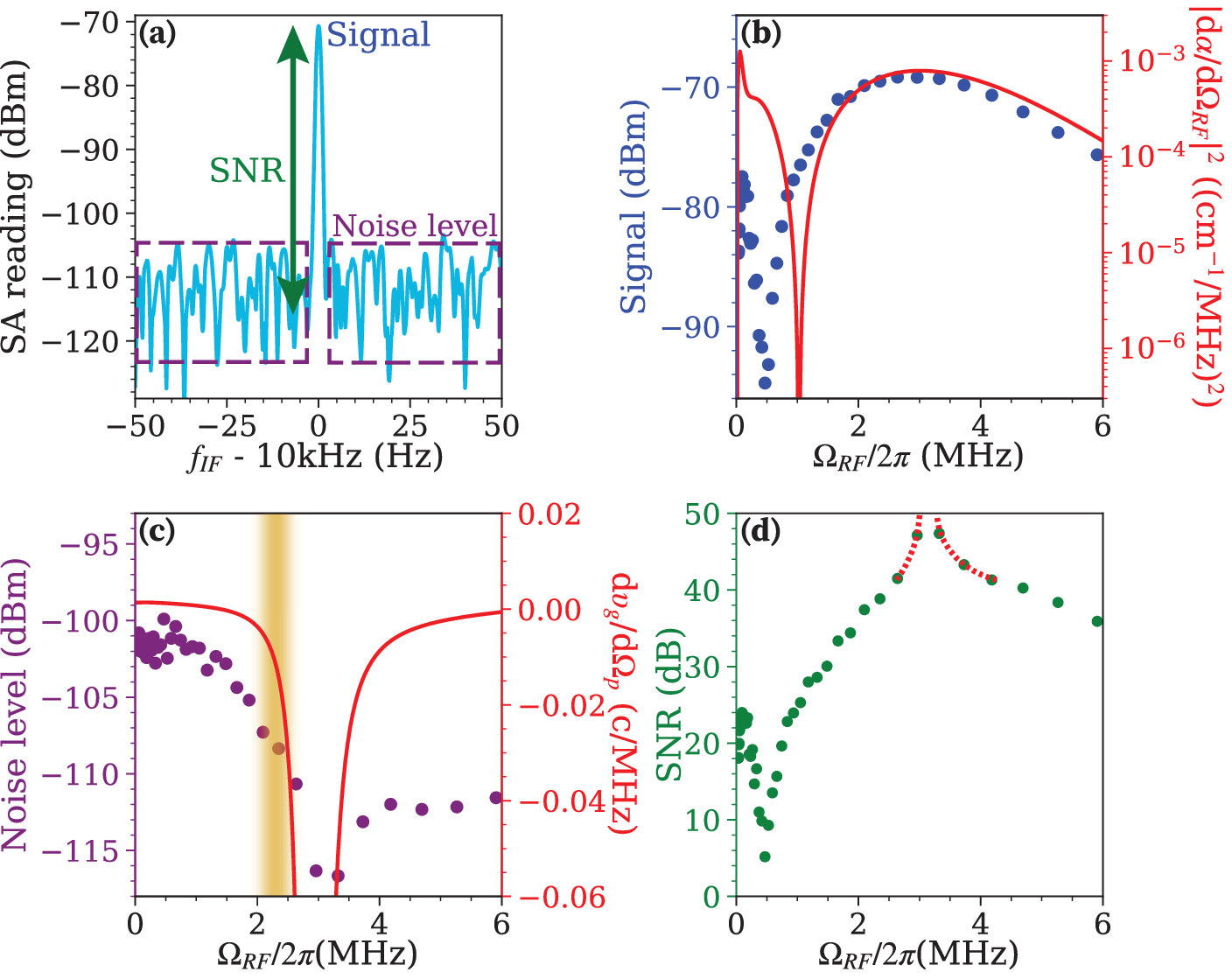}
\caption{(a) SA spectrum of the Rydberg-receiver output signals for a measurement of  9.937~GHz fields at LO and SIG electric fields of $E_{LO}=$~0.2~V/m and $E_{SIG}=$~0.53~mV/m, respectively.
(b) Theoretical responsivity squared, $|d\alpha / d\Omega_{RF}|^2$, as a function of $\Omega_{RF}$ overlaid with the measured heterodyne signal at IF~=~10~kHz (blue dots). (c) Calculated $dv_g/d\nu_p$ (solid red line) overlaid with the measured SA noise level (purple symbols) at IF~=~10~kHz. Note the noise dip centered at $\Omega_{RF}/(2\pi) \approx 3.1$~MHz. The orange shading indicates the region where negative feedback in the calculation suppresses polariton noise (see text for details). (d) Measured SNR as a function of $\Omega_{RF}$. The red dotted curve is a visual guide that highlights the SNR enhancement afforded by polariton-noise tuning.}
\label{fig:snr}
\end{figure}

In Fig.~\ref{fig:snr}, we provide an example of Rydberg-HET signal and noise measurements. The probe power, $P_{out}(t)$, is analyzed on the SA, revealing the beat signal at the selected $f_{IF}$ of 10~kHz [labeled ``Signal" in Fig.~\ref{fig:snr}~(a)]. The accompanying noise is outlined by dashed rectangles and labeled ``Noise level" in Fig.~\ref{fig:snr}~(a). Here, a resolution bandwidth of 1~Hz is used. The signal level (blue dots) is then plotted in Fig.~\ref{fig:snr}~(b) as a function of the utilized LO Rabi frequency, $\Omega_{RF}$. The optimum signal is observed near $\Omega_{RF}/(2\pi) = 3$~MHz. The theoretical response scales with $| d \alpha / d \Omega_{RF}|^2$, the curve that is overlaid with the experimental data [red line in Fig.~\ref{fig:snr}~(b)]. We see that 30~dB in measured gain corresponds to a change in $| d \alpha / d \Omega_{RF}|^2$ by a factor of approximately 35~dB [note the span of the right red-colored ordinate axis in Fig.~\ref{fig:snr}~(b)]. Given the complexity of the physical system and the underlying calculations, the measured and calculated responses exhibit good overall agreement, lending confidence in the validity of the model. Residual discrepancies at $\Omega_{RF}/(2\pi) < 1$~MHz may arise from RF-field inhomogeneities and nonlinear effects in Rydberg HET when SIG and LO field amplitudes become comparable~\cite{luis2024nonlinearity}. 

In Fig.~\ref{fig:snr}~(c), we show the SA noise level as a function of $\Omega_{RF}$. At $\Omega_{RF}$ near zero, polariton noise adds up to $\sim 20$~dB of noise above the optical noise (also see Table~\ref{tab:tab1}). The noise floor drops by $\sim 15$~dB at $\Omega_{RF}/(2\pi) \approx 3.1$~MHz. Finally, in Fig.~\ref{fig:snr}~(d), we plot the extracted SNR as a function of $\Omega_{RF}$. There, it is observed that a choice of the LO Rabi frequency at $\Omega_{RF}/(2\pi) = 3.1$~MHz (with an uncertainty of $\lesssim 100$~kHz) allows us to reduce the polariton noise by $\gtrsim 5$~dB. At $\Omega_{RF}/(2\pi) = 3.1$~MHz, this leaves the remaining polariton noise of only about 5~dB above the optical noise. This improvement is afforded by the fact that the noise minimum [Fig.~\ref{fig:snr}~(c)] and the signal maximum [Fig.~\ref{fig:snr}~(b)] coincide at the chosen LO operating point of $\Omega_{RF}/(2\pi) = 3.1$~MHz.

\section{Discussion and Conclusion}
\label{discussion}

In this work, we showed that Rydberg EIT in thermal vapors is accompanied by an increased noise level in the detected probe light that is attributed to the propagation of Rydberg polaritons through the EIT medium. This effect can increase the overall noise by approximately 20-25~dB under typical operating conditions. This limits the sensitivity of Rydberg electrometers and negatively affects the performance of Rydberg-HET receivers. At the same time, we found that the RF field used for coherent RF readout in Rydberg-HET receivers also reshapes the polariton propagation dynamics and can tune the medium into a low-noise operating point. This establishes the reference RF field as an active control knob for both signal generation and polariton-noise suppression. The resulting SNR enhancement is achieved by tuning the reference-field Rabi frequency into a local minimum of the polariton-noise landscape that nearly coincides with the signal maximum. In our demonstration, the $\Omega_{RF}$-value is controlled at a level of about 3\% to harvest the SNR enhancement afforded by polariton noise control. While our work is conducted using room-temperature atomic vapor, it is noteworthy that the storage and control of Rydberg polaritons using RF fields has also been demonstrated in a cold-atom medium~\cite{adams2013control}.

Our findings are supported by a physical model that reveals the existence of conditions under which the Rydberg-EIT system operates in the negative feedback regime. There, the polariton medium serves as a ``noise buffer'' for fluctuations of the transmitted probe power. The demonstrated polariton noise reduction exploits the system's ability to retain and release probe-beam energy in a way that reduces fluctuations of the transmitted probe power.

We note that the probe's group velocity depends on both probe and coupler Rabi frequencies. Therefore, a comprehensive investigation involving fine-tuning of relevant parameters (powers, beam diameters and profiles) as well as time-dependent studies~\cite{Sapiro_2020, bohaichuk2022} can be done in the future. Given that the polariton propagation time through the Rydberg-EIT medium is on the order of $L/v_g\approx 100$~ns, we expect that the discussed feedback between probe-beam power and group velocity will generate (anti-)correlations in the probe-photon counting statistics. Furthermore, such measurements may reveal additional insights into polariton propagation within the EIT medium~\cite{Firstenberg2013, Cantu2020}. These aspects would be of interest in the field of quantum optics and quantum information science, including the generation of non-classical states of light~\cite{Eisaman2005, Neuzner2016}, quantum memories and repeaters~\cite{Chaneliere2005, Katz2018}, and optical squeezing~\cite{jinprl2024}.

\begin{acknowledgments}
This work was supported by Rydberg Technologies Inc. and, in part, by the Defense Advanced Research Projects Agency (DARPA) SAVaNT Program under agreement No. HR00112190065. Distribution Statement ``A" (Approved for Public Release, Distribution Unlimited).  The views and conclusions contained in this document are those of the authors and should not be interpreted as representing the official policies, either expressed or implied, of the U.S. Government.
\end{acknowledgments}

\bibliography{bibfile}

\end{document}